# MARS: Feature-guided physics-informed deep learning for real-time battery health and lifetime prediction

Innocent Izabayo *Graduate Student Member, IEEE*, Haiquan Zhao, *Senior Member, IEEE*

***Abstract*— Accurate and robust estimation of battery state-of-health (SOH) and remaining useful life (RUL) is essential for safe and cost-effective operation of electrified transportation systems, yet it remains challenging under operating variability, non-stationary aging, and non-Gaussian measurement disturbances. This paper proposes Multi-scale Adaptive Robust State Estimation (MARS), a multi-scale adaptive robust state-estimation framework that couples a dual-branch temporal encoder with an equivalent-circuit-model-constrained differentiable generalized-correntropy adaptive Kalman filter. The encoder disentangles fast intra-cycle voltage-current-temperature dynamics from slow inter-cycle degradation, and the fused degradation context adaptively parameterizes the cycle-level process covariance, measurement covariance, and correntropy bandwidth. Through the unrolled robust filtering recursion, MARS performs physics-consistent state propagation and generalized-correntropy-based innovation correction, while uncertainty decomposition and Monte Carlo dropout provide calibrated prediction intervals for SOH estimation and conditional RUL prediction. Experiments on the NASA and CALCE benchmarks demonstrate improved SOH accuracy, achieving mean absolute error/root mean square error values of 0.72%/0.92% and 0.63%/0.80%, respectively, together with competitive conditional RUL performance against representative data-driven and hybrid baselines. The resulting conditional-RUL prediction intervals achieve coverage probabilities of 100.00% and 99.24% on NASA and CALCE, respectively. Ablation, disturbance-sensitivity, near-end-of-life, and runtime analyses further verify the contribution of each module and the practical inference cost for online battery management.***

## I. INTRODUCTION

Lithium-ion batteries (LIBs) in electrified vehicles (EVs) are safety- and performance-critical subsystems; their aging directly affects usable range, charging strategy, availability, and total cost of ownership [1], [2]. During service, coupled electrochemical and thermal degradation reduces usable capacity and power capability and can increase operational risk when health is not monitored and managed [3], [4], [5]. In practice, state of health (SOH) is commonly quantified using capacity retention relative to the initial rated value, and end-of-life is often associated with a specified retention band [6], [7], [8]. Accurate SOH estimation and remaining useful life (RUL) prediction are therefore enabling functions for battery management systems (BMS), supporting preventive maintenance, warranty-risk management, and fleet-level planning [9].

Reliable SOH and RUL prognostics remain difficult in vehicle settings because battery aging is inherently multi-timescale. Fast within-cycle dynamics captured in voltage, current, and temperature trajectories coexist with slow inter-cycle aging evolution, and both are modulated by operating variability such as temperature variation, charge/discharge rates, and partial cycling [10], [11]. This diversity introduces non-stationarity and distribution shift across cells and duty cycles, while sensing artifacts and disturbances may be heavy tailed rather than Gaussian. For transportation-grade deployment, prognostic outputs must remain physically plausible and robust under non-Gaussian disturbances while providing uncertainty information at bounded latency and memory cost [12]-[14]. In real-time vehicle environments, sensing noise often deviates from ideal Gaussian white noise because of electromagnetic interference from inverter switching and regenerative braking, vibration-induced variations in connector contact resistance, sensor quantization, and rapid temperature or load transients that introduce structural mismatch in the equivalent-circuit model (ECM) [15].

Model-based strategies (e.g., empirical degradation laws and equivalent-circuit models) are attractive for online use because they are interpretable, but they often require retuning across aging stages and operating regimes, limiting transferability [5]. Sequential Bayesian estimators can represent nonlinearities and uncertainty, yet they are sensitive to model mismatch and tuning choices and can incur nontrivial computational overhead on embedded hardware [16]. Electrochemical models provide strong mechanistic grounding, but their calibration burden and runtime cost still make real-time, on-board deployment challenging in many applications [17]. These limitations have encouraged hybrid estimators that combine learned models with physically interpretable estimation methods [15]. For example,

This work was partially supported by National Natural Science Foundation of China (grant: 62671522, 62171388, 61871461).
Izabayo Innocent, Haiquan Zhao are with the key Laboratory of Magnetic Suspension Technology and Maglev Vehicle, Ministry of Education, School of Electrical Engineering Southwest Jiaotong University Chengdu, China. (e-mail: izabayo37@gmail.com; hqzhao_swjtu@126.com)
* Corresponding author: Haiquan Zhao

a recent vehicle longitudinal-dynamics study coupled a neural traction model with recursive least-squares estimation to identify parameters such as vehicle mass and road friction from limited driving data [18]. Furthermore, managing adaptive state estimation in real-time BMS applications requires balancing parameter tracking accuracy against processing overhead under transient driving profiles, a trade-off that has been addressed in adaptive state of power (SOP) estimation using fuzzy driving pattern recognition and multi-criteria decision-making [19].

Data-driven methods can learn nonlinear relationships between measured signals or engineered indicators and health targets. Convolutional and recurrent architectures have been extensively used to extract temporal degradation signatures and learn time-dependent aging behavior from battery data [14], [20], [21], [22], and hybrid deep structures that combine complementary temporal modules continue to improve RUL prediction performance in transportation-relevant studies [23]. Attention-based models have also gained traction because they can represent long-range dependencies more directly, enabling both early-life inference and long-horizon forecasting [8], [24]. Feature engineering remains an effective complement, and extracting health factors from cycling data has been shown to improve estimation accuracy and efficiency [7]. Nevertheless, purely data-driven predictors may generate physically implausible extrapolations under operating shifts and often provide limited or poorly calibrated uncertainty characterization, which constrains their utility for safety-critical decisions [6], [12]. Hybrid and physics-informed methods attempt to constrain predictions to follow degradation structure and operating limits [25], including cloud-edge schemes for resource-limited deployment [21], physics-regularized learning and physics-informed neural networks that embed degradation structure such as monotonic behavior [17], [26], and multi-scale learning to better describe heterogeneous aging dynamics [9]. Overall, these elements are rarely combined end-to-end for SOH estimation and conditional RUL prognostics, particularly when robust state propagation under non-Gaussian disturbances and predictive uncertainty are both required [12].

Robust filtering under non-Gaussian noise is particularly relevant to transportation sensing environments [27]. Generalized correntropy provides a robust alternative to mean-squared objectives in the presence of outliers and heavy-tailed disturbances [28], [29], [30]. Correntropy-based Kalman variants have improved robustness in non-Gaussian settings [31], and battery studies have reported gains when generalized correntropy is incorporated into filtering [32]. Pipelined nonlinear and recurrent filters have been investigated for dynamic system identification [33]-[35], memory-proportionate affine projection methods for sparse echo-path tracking [36], robust data-reuse adaptive filtering for noisy-input conditions [37], and correntropy-based unscented Kalman filters for non-Gaussian state estimation [38], [39]. Nevertheless, existing robust Kalman approaches typically operate at a single timescale, rely on heuristic adaptation rules, and are seldom trained end-to-end together with multi-timescale feature extractors and uncertainty-quantification mechanisms.

Motivated by these gaps, this paper proposes MARS (Multi-scale Adaptive Robust State Estimation), a feature-guided framework for SOH estimation and conditional RUL prediction with quantified uncertainty. MARS integrates a dual-branch temporal encoder with a differentiable generalized-correntropy adaptive Kalman filter whose process covariance, measurement covariance, and correntropy scale are conditioned on the learned degradation context. This coupling enables physics-consistent state propagation, robust innovation correction, and uncertainty-aware prediction. The framework is evaluated on the NASA and CALCE benchmarks under distinct degradation horizons and cell-aging patterns. The main contributions are summarized as follows:

1. We propose MARS, a multi-scale adaptive robust state-estimation framework for lithium-ion battery SOH estimation and conditional RUL prediction. It integrates dual-branch temporal encoding, ECM-constrained robust filtering, and uncertainty-aware prediction within a unified end-to-end prognostic pipeline.
2. We develop a tightly coupled differentiable generalized-correntropy adaptive Kalman filter that uses the fused degradation context to adapt the process covariance, measurement covariance, and correntropy bandwidth at each cycle. This enables physics-consistent state propagation and robust innovation correction under heavy-tailed disturbances, rather than relying on fixed or offline-tuned filtering statistics.
3. We develop an uncertainty-aware SOH/RUL prognostic architecture that fuses the multi-scale degradation representation with the filtered posterior state through task-specific prediction heads. The predictive uncertainty is decomposed into aleatoric, epistemic, and state-covariance-derived components to generate calibrated point estimates and prediction intervals.

The remainder of this paper is organized as follows. Section II describes the datasets, preprocessing, and label construction. Section III presents the proposed MARS methodology, including the multi-scale encoder, differentiable robust filtering, and uncertainty-aware prediction. Section IV reports the experimental settings and evaluation results. Section V concludes the paper.

## II. Datasets and Preprocessing

We evaluate MARS on two public lithium-ion battery repositories with distinct cell formats and test protocols: the NASA Ames dataset [40] and the CALCE database [12]. The NASA subset contains commercial cylindrical 18650 NCA cells rated at 2.0 Ah, with end-of-life (EOL) defined at 70% capacity retention (1.4 Ah). The CALCE subset contains prismatic LCO cells rated at 1.1 Ah, with EOL defined at 80% capacity retention (0.88 Ah) to match standard benchmark conventions.

These datasets provide complementary degradation regimes that are relevant to transportation-battery prognostics, although they remain cell-level benchmark datasets rather than full vehicle-pack experiments. The NASA cells exhibit short lifetimes (~ 150 cycles) and frequent capacity-regeneration events, reflecting short-horizon capacity-regeneration behavior that is relevant to intermittent-use battery operation. Conversely, the CALCE cells provide longer trajectories (~ 700-800 cycles) with gradual capacity fade and a distinct late-life degradation knee, providing a longer-horizon degradation pattern relevant to continuous service operation. These contrasting regimes evaluate whether MARS can represent both rapid within-cycle transients and slowly-evolving inter-cycle capacity fade while maintaining stable state propagation near the EOL.

During service, voltage, current, and temperature trajectories form the short-horizon inputs, whereas cycle-level descriptors and their temporal increments characterize long-term degradation. Incremental-capacity $(dQ/dV)$ signatures are extracted as degradation-sensitive health factors to capture aging-induced electrochemical changes [7], and the OCV-SoC relationship supports the GCAKF measurement model. Furthermore, the unrolled state-space formulation in Eq. (10) represents equivalent-circuit states and parameters, including resistance, capacity, and polarization, that are broadly applicable to common automotive battery chemistries and cell formats. Fig. 1 summarizes the capacity trajectories, SOH distributions, cycle-to-cycle variations, RUL profiles, incremental-capacity signatures, and OCV-SoC characteristics across both benchmarks.

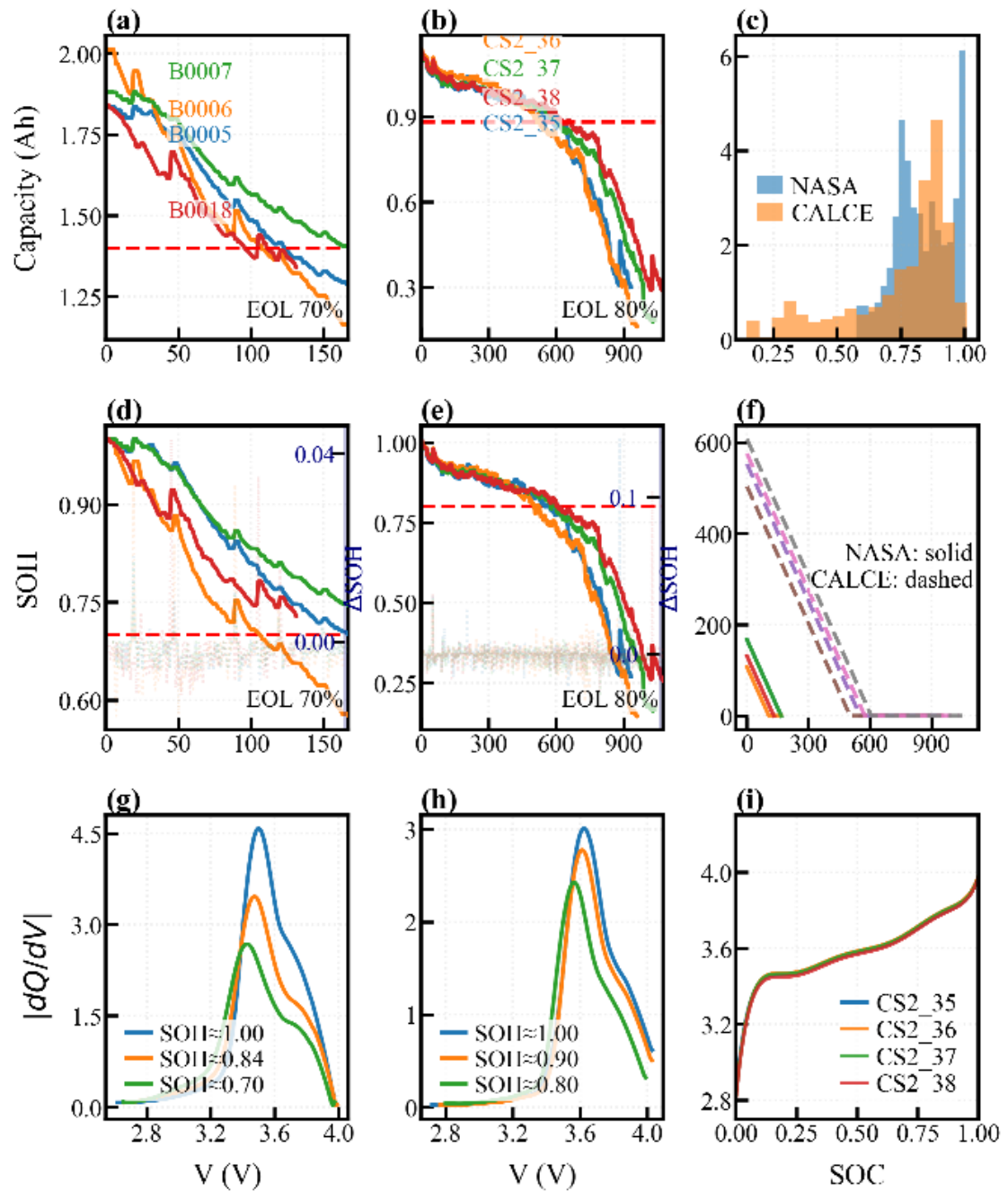

*Fig. 1. Dataset overview of NASA and CALCE cells. (a)-(b) capacity fade; (c) SOH distributions; (d)-(e) SOH with* $\Delta$SOH; *(f) RUL trajectories; (g)-(h)* $|dQ/dV|$ *signatures; (i) OCV curves.*

*1)Label construction.* For each cycle *k*, we isolate the primary discharge segment and compute the discharge capacity $Q_k$ via trapezoidal integration of the measured current trajectory (Coulomb counting). Because capacity fade is the standard operational proxy for loss of usable lithium inventory and active material under a fixed protocol, we define:

$$\mathrm{SOH}_k = \frac{Q_k}{Q_0} \in [0,1], \tag{1}$$

where $Q_0$ is the beginning-of-life capacity estimated from early-life cycles of dataset *d*. End-of-life (EOL) is defined using a dataset-dependent capacity-retention threshold $\delta_d$:

$$k_{EOL}^{(d)} \triangleq \min\{k : \mathrm{SOH}_k^{(d)} \le \delta_d\} \tag{2}$$

with $\delta_{NASA} = 0.70$ and $\delta_{CALCE} = 0.80$ to match the respective benchmark conventions. The remaining useful life (RUL) label is computed as.

$$\mathrm{RUL}_k^{(d)} \triangleq \max\left(0, k_{EOL}^{(d)} - k\right). \tag{3}$$

If RUL is learned in a normalized space, the mapping back to actual cycle counts is used only for evaluation and reporting.

*2)Multi-scale cycle representation*. Each cycle is represented at two complementary timescales to capture short-horizon electrochemical dynamics and long-horizon degradation context. The short-horizon input uses the discharge voltage, current, and temperature sequence and is mapped to a fixed length (L=50) by truncation or right-padding to enable batching. In parallel, a cycle-level engineered descriptor adopts a state-delta design, yielding a 50-dimensional feature vector that concatenates a cycle state and its cycle-to-cycle increment. Incremental-capacity $(dQ/dV)$ characteristics serve as informative aging proxies (Fig. 1(g)-(h)), and an OCV-SoC mapping is constructed to support the physics-based state evolution component (Fig. 1(i)).

*3)Multi-Scale Feature Construction.* All features are normalized using min-max scaling fitted on the training set and applied to validation/test sets. We use a fixed random seed (seed=42) and a 3:1:1 split for training/validation/testing, and all baselines and ablations follow identical data definitions for fair comparison.

## III. Proposed Methodology

Practical lithium-ion battery prognostics must remain numerically stable and bias-resistant under nonstationary aging, including knee-point acceleration, and heavy-tailed measurement disturbances arising from intermittent outliers, bias drift, and sensor artifacts. Accordingly, MARS couples learned degradation context with a physics-constrained, robustness-weighted Kalman recursion to mitigate innovation-driven drift over the lifecycle. MARS (Multi-scale Adaptive Robust State Estimation) integrates a dual-scale Transformer encoder that disentangles intra-cycle transients from inter-cycle degradation dynamics, a differentiable generalized-correntropy adaptive Kalman filter (GCAKF) that enforces physically consistent state evolution while rejecting heavy-tailed innovations, and an uncertainty-aware multi-task predictor for joint SOH and RUL inference with calibrated predictive intervals. Fig. 2 summarizes the information flow and the closed-loop interaction between degradation representations and physics-based filtering.

### *A. Problem formulation and multi-scale inputs*

We index cycles by $c \in \{1,..,C\}$ and denote within-cycle samples by $k \in \{1,..,L\}$ after resampling to a uniform sampling interval $\Delta t = 1s$ to align electrochemical transients across heterogeneous datasets. At cycle $c$, MARS consumes a short-horizon discharge segment to capture intra-cycle electrochemical transients, together with a long-horizon sequence of engineered cycle descriptors to capture slow degradation evolution. These two inputs are encoded into a degradation context vector that conditions the subsequent robust state-estimation recursion and multi-task prediction.

The short-horizon input at cycle $c$ is the resampled discharge segment $\{[V_{c,k}, I_{c,k}, T_{c,k}]\}_{k=1}^{L}$, arranged as

$$X_c^{short} = \begin{bmatrix} V_{c,1} & \cdots & V_{c,L} \\ I_{c,1} & \cdots & I_{c,L} \\ T_{c,1} & \cdots & T_{c,L} \end{bmatrix} \in R^{3\times L}. \quad (4)$$

The long-horizon input is a window of engineered cycle descriptors over the most recent $W$ cycles,

$$X_c^{long} = \begin{bmatrix} f_{c-W+1} \\ \vdots \\ f_c \end{bmatrix} \in R^{W\times D_{long}}, \quad (5)$$

where $f_c \in R^{D_{long}}$ denotes the engineered descriptor vector for cycle $c$, and $D_{long}$ is the descriptor dimensionality.

The supervised targets are SOH and RUL, defined in Section II. When SOH and RUL are co-estimated, SOH, which represents capacity retention, provides the current health-state boundary condition for the forward-looking RUL trajectory [15]. Because future load and environmental variations introduce epistemic uncertainty during forecasting, co-estimating the current health state helps stabilize the long-term degradation trend and prevents non-physical prognostic projections, such as predicting a long remaining life when the SOH is already close to the EOL threshold. When the model is trained with normalized RUL labels, the predicted RUL is converted back to actual cycle counts only for evaluation and reporting.

### B. Dual-scale Transformer encoder and fusion

Panel B in Fig. 2 encodes the two streams using a short-term and a long-term Transformer to obtain complementary representations:

$$\boldsymbol{h}_k^{(s)} = \mathrm{Enc}_{short}(\boldsymbol{X}_k^{short}), \quad \boldsymbol{h}_k^{(\ell)} = \mathrm{Enc}_{long}(\boldsymbol{X}_k^{long}), \quad (6)$$

which are fused by a multi-layer perceptron (MLP) with a sigmoid gating layer to yield a compact, cycle-level context:

$$\boldsymbol{h}_{f,k} = \boldsymbol{W}_g \cdot \left[ \boldsymbol{h}_k^{(s)}; \boldsymbol{h}_k^{(l)} \right] \odot \sigma\left(\boldsymbol{W}_c \cdot \boldsymbol{h}_k^{(l)} + \boldsymbol{b}_c\right) \quad (7)$$

where $\boldsymbol{W}_g$ and $\boldsymbol{W}_c$ are learned projection matrices, $\boldsymbol{b}_c$ is a bias vector, and $\odot$ is the Hadamard product. This formulation ensures that the slow degradation trend dynamically modulates the sensitivity of the output to fast transients. The baseline values reported in Table I are derived from offline electrochemical parameter characterization of 18650 and prismatic cells [15] to serve as robust physical priors, which are then adaptively scaled during training via the feature-conditioned scaling factor $s_{f,k}$ to maintain stable initialization bounds.

Transformers are adopted due to their self-attention mechanism, which supports global dependency modeling in long, non-stationary sequences and is well-suited to battery time-series where relevant cues may be distributed across time [12], [41].

*1)De-stationary long-term encoding.* Because degradation trajectories are typically non-stationary, we separate a window trend from residual variations before attention. Let $\boldsymbol{Z}_k^{(\ell)} \in R^{W\times d_\ell}$ denote long-window token embeddings. We compute a window mean:

$$\boldsymbol{\tau}_k = \frac{1}{W}\sum_{j=1}^{W} \boldsymbol{Z}_k^{(\ell)}(j,:), \quad \boldsymbol{S}_k(j,:) = \boldsymbol{Z}_k^{(\ell)}(j,:) - \boldsymbol{\tau}_k, \quad (8)$$

and refine the sequence through a residual projection:

$$\boldsymbol{Z}_k^{(\ell)'} = \boldsymbol{Z}_k^{(\ell)} + \boldsymbol{W}_d\left[\boldsymbol{1\tau}_k; \boldsymbol{S}_k\right], \quad (9)$$

followed by the long-term Transformer and pooling to yield. Here $\boldsymbol{1\tau}_k$ denotes replication of $\boldsymbol{\tau}_k$ across the W tokens.

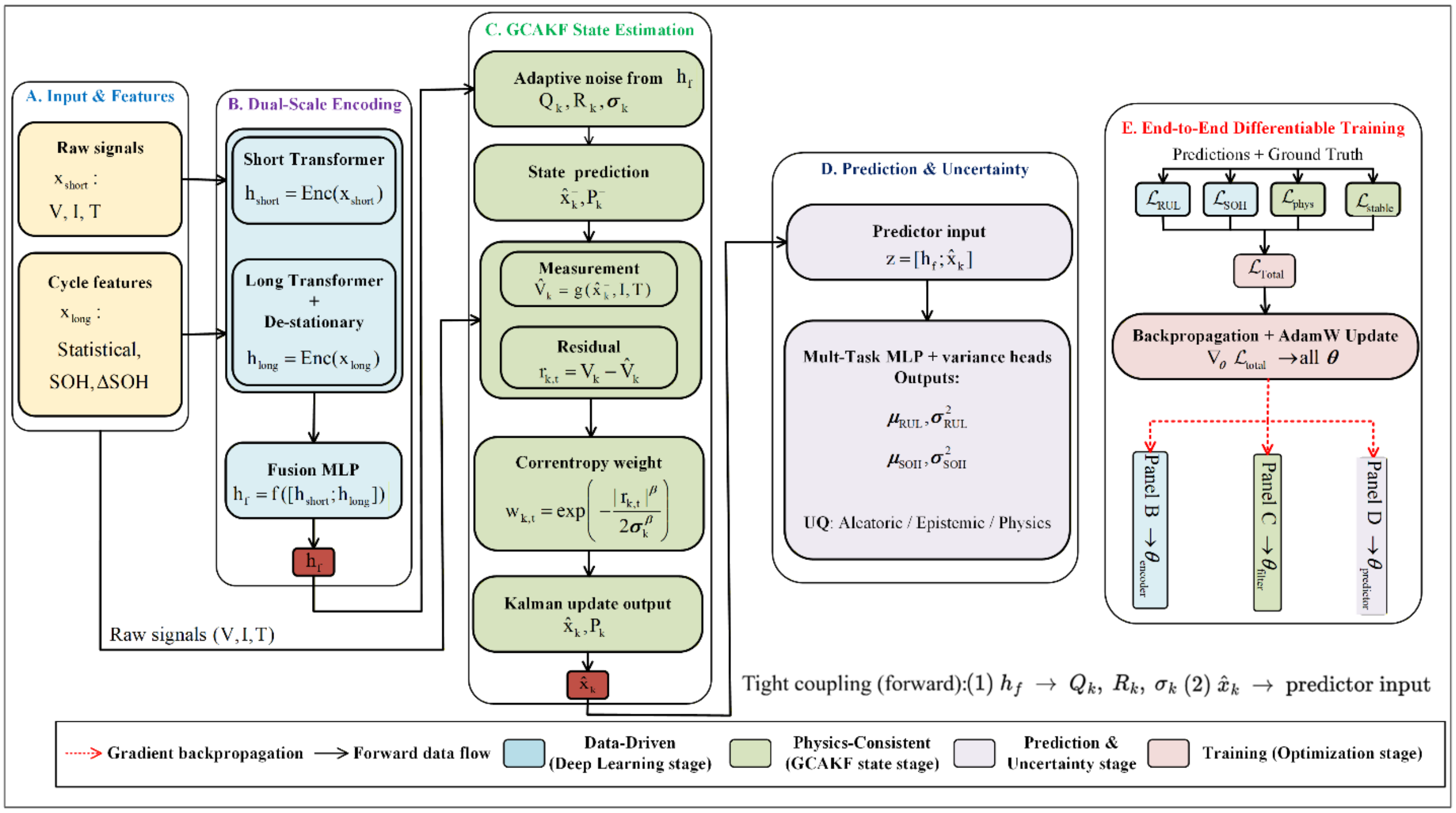

*Fig. 2. Overview of the proposed MARS framework: dual-scale encoding (Panel B), differentiable GCAKF state estimation (Panel C), multi-task prediction with uncertainty (Panel D), and end-to-end training via backpropagation (Panel E).*

*C. Differentiable robust state estimation (GCAKF)*

Panel C of Fig. 2 performs physics-informed state estimation via a differentiable GCAKF built on first-order RC equivalent-circuit model (Thevenin) illustrated in Fig. 3. The latent state aggregates physically meaningful variables required for downstream inference:

$$\boldsymbol{x}_{k,t} = \left[ SoC_{k,t}, U_{1,k,t}, Q_{max,k,t}, R_{0,k,t}, R_{1,k,t}, C_{1,k,t} \right]. \tag{10}$$

The latent state vector aggregates physically meaningful variables $\boldsymbol{x}_{k,t} = [SoC, U_1, Q_{max}, R_0, R_1, C_1]$ (Fig. 3). To ensure reproducible and stable convergence within the unrolled differentiable framework, the GCAKF parameters are initialized with the baseline values reported in Table I. These initializations reflect the relative uncertainty of slowly varying physical dynamics compared to fast states. During the training phase, these parameters are adaptively scaled by the feature-conditioned module conditioned on $\boldsymbol{h}_{f,k}$, as in Section III-C.2.

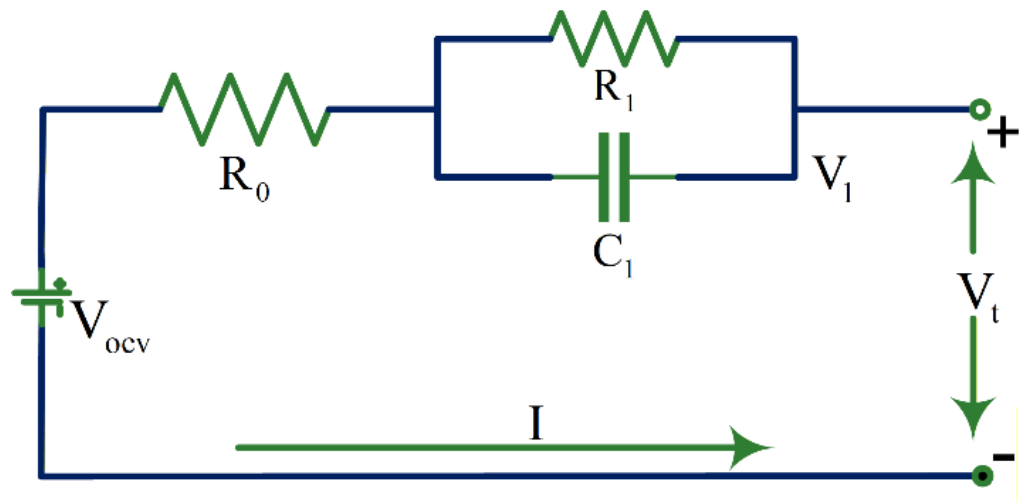

*Fig. 3. First-order RC equivalent-circuit model.*

*1)ECM dynamics and measurement model.* Consistent with Fig. 3, the discretized within-cycle dynamics (sampling interval $\Delta t$) are

$$\begin{aligned} SoC_{k,t} &= SoC_{k,t-1} - \frac{\eta \Delta t}{Q_{max,k,t-1}} I_{k,t-1}, \\ U_{1,k,t} &= a_{k,t} U_{1,k,t-1} + (1-a_{k,t}) R_{1,k,t-1} I_{k,t-1}, \\ a_{k,t} &= \exp\left( -\frac{\Delta t}{R_{1,k,t-1} C_{1,k,t-1}} \right). \end{aligned} \tag{11}$$

while slowly varying parameters (e.g., $Q_{max}, R_0, R_1, C_1$) are modeled as random walks. The terminal-voltage measurement model is

$$V_{k,t} = OCV(SoC_{k,t}, T_{k,t}) - I_{k,t} R_{0,k,t} - U_{1,k,t} + v_{k,t}, \tag{12}$$

where $v_{k,t}$ is measurement noise and $OCV(\cdot)$ may be temperature-dependent [25].

*2)Feature-conditioned uncertainty and robustness.* A key distinction from classical EKF practice is that the cycle-level process covariance $\boldsymbol{Q}_k$, measurement covariance $\boldsymbol{R}_k$, and correntropy robustness scale $\sigma_k$ are learned as cycle-dependent quantities conditioned on $\boldsymbol{h}_{f,k}$ Fig. 2 B to C:

$$\boldsymbol{Q}_k = \exp(s_{T,k}) \exp(s_{f,k}) \boldsymbol{\Lambda}_Q, \quad \boldsymbol{R}_k = \exp(s_{T,k}) \exp(s_{f,k}) \boldsymbol{\Lambda}_R, \tag{13}$$

$$\sigma_k = \mathrm{softplus}(\tilde{\sigma}_k) \tag{14}$$

where $\mathrm{softplus}(x) = \ln(1+\exp(x))$ ensures a strictly positive scale mapping, $\boldsymbol{\Lambda}_Q$ and $\boldsymbol{\Lambda}_R$ represent diagonal baseline covariance matrices derived from offline electrochemical calibration [15], $s_{T,k} \in \mathbb{R}$ is a temperature-dependent scaling scalar, and $s_{f,k} \in \mathbb{R}$ is a feature-conditioned scaling factor predicted from the context vector $\boldsymbol{h}_{f,k}$.

*3) Robust measurement update via generalized correntropy*. For each discharge sample $t = 1,..,L$, the GCAKF suppresses heavy-tailed innova-tions using a generalized correntropy [28], [29]. The innovation is $r_{k,t} = V_{k,t} - \hat{V}_{k,t}$, where $\hat{V}_{k,t}$ is the terminal voltage predicted by the ECM in Eq. (11).

The correntropy weight is

$$w_{k,t} = \exp\left( -\frac{|r_{k,t}|^{\beta}}{2\sigma_k^{\beta}} \right), \qquad 0 < \beta \leq 2, \tag{15}$$

where $\sigma_k$ is the adaptive kernel bandwidth and $\beta$ is the shape parameter. We restrict the shape parameter to $0 < \boldsymbol{\beta} \leq 2$ based on Mercer kernel constraints and robust M-estimation principles under heavy-tailed noise [29], [42]. The generalized Gaussian kernel is a positive-definite Mercer kernel if and only if $\boldsymbol{\beta} \leq 2$, which is required to guarantee stable state projections in the Reproducing Kernel Hilbert Space (RKHS) [42]. While values of $\boldsymbol{\beta} > 2$ can accelerate convergence under light-tailed sub-Gaussian noise, they do not yield a redescending influence function [42]. Setting $\boldsymbol{\beta} \leq 2$ ensures that the measurement update gain decays asymptotically to zero for large residuals, suppressing impulsive outliers and sensor spikes [29]. Robustness is enforced by inflating the effective measurement covariance:

$$\widetilde{\boldsymbol{R}}_{k,t} = \frac{\boldsymbol{R}_k}{w_{k,t} + \varepsilon}, \tag{16}$$

with $\varepsilon = 10^{-3}$ ensuring numerical stability. To maintain numerical stability and preserve the positive semi-definiteness of the covariance under finite-precision backpropagation, we apply the Joseph stabilized covariance update:

$$\boldsymbol{P}_{k,t} = (\boldsymbol{I} - \boldsymbol{K}_{k,t}\boldsymbol{H}_{k,t})\boldsymbol{P}_{k,t}^{-}(\boldsymbol{I} - \boldsymbol{K}_{k,t}\boldsymbol{H}_{k,t}) + \boldsymbol{K}_{k,t}\widetilde{\boldsymbol{R}}_{k,t}\boldsymbol{K}_{k,t}. \tag{17}$$

As $|r_{k,t}|$ increases, the exponential decay of $w_{k,t}$ increases $\widetilde{\boldsymbol{R}}_{k,t}$ and suppresses the Kalman gain $\boldsymbol{K}_{k,t}$, bounding the influence of non-Gaussian outliers on the posterior $(\hat{x}_{k,t}, \boldsymbol{P}_{k,t})$.

The terminal state estimate is taken as the end-of-cycle posterior:

$$(\hat{x}_k, \boldsymbol{P}_k) \triangleq (\hat{x}_{k,L}, \boldsymbol{P}_{k,L}), \tag{18}$$

which is forwarded for joint prognostic inference.

*D. Multi-task prediction and uncertainty decomposition*

As illustrated in Panel D in Fig. 2 MARS couples the learned temporal context with the physics-consistent state estimate by forming a joint feature vector:

$$\boldsymbol{z}_k \triangleq \left[ \boldsymbol{h}_{f,k}; \hat{\boldsymbol{x}}_k \right] \tag{19}$$

where $\hat{\boldsymbol{x}}_k$ is the end-of-cycle posterior derived from Eq. (18). A multi-task MLP with task-specific heads then outputs point estimates and aleatoric variances $(\mu_{y,k}, \sigma_{ale,y}^2(k))$ for $y \in \{RUL, SOH\}$. To ensure numerical stability, the aleatoric variance is bounded via:

$$\sigma_{ale,y}^2(k) = \sigma_{min,y}^2 + \left( \sigma_{max,y}^2 - \sigma_{min,y}^2 \right) \mathrm{sigmoid}(s_{y,k}) \tag{20}$$

where $s_{y,k}$ denotes the raw output of the corresponding variance head.

To support risk-aware battery management [12], the total predictive variance is decomposed into aleatoric, epistemic (MC-dropout), and physics-induced components:

$$\sigma^2_{\text{tot},y}(k) = \sigma^2_{\text{ale},y}(k) + \sigma^2_{\text{epi},y}(k) + \sigma^2_{\text{phys}}(k) \qquad (21)$$

Epistemic uncertainty $\sigma^2_{\text{epi},y}(k)$ is approximated using MC dropout with $M$ stochastic forward passes to capture the model-weight distribution. Physics-induced uncertainty propagates state-estimation ambiguity from the GCAKF posterior covariance $\boldsymbol{P}_k$ using a normalized scalar proxy:

$$u_{\text{phys}}(k) = \sqrt{\text{mean}\left(\text{diag}(\boldsymbol{P}_k) \oslash \boldsymbol{s}^2\right)}, \quad \sigma^2_{\text{phys}}(k) = u^2_{\text{phys}}(k) \qquad (22)$$

where $\oslash$ denotes element-wise division and $\boldsymbol{s}$ is a fixed nominal state-scale vector. The unrolled GCAKF covariance $\boldsymbol{P}_k$ (Eq. 17) encodes physical state uncertainty under operating variability, capturing the estimation error covariance of crucial capacity and resistance parameters. Under sensor noise or model mismatch, GCAKF updates naturally inflate $\boldsymbol{P}_k$ through innovation residual scaling. Through Eq. (22), its diagonal elements are mapped to the physics-induced uncertainty proxy $\boldsymbol{\sigma}^2_{\text{phys},y}(k)$ for the target $y \in \{\text{SOH}, \text{RUL}\}$ and incorporated into the total variance $\boldsymbol{\sigma}^2_{\text{tot},y}(k)$, allowing prediction intervals to expand during transient, noisy, or poorly modeled cycles and contract when degradation is stable. This physical regularization of uncertainty boundaries aligns with advanced physics-informed frameworks leveraging physical constraints and attention mechanisms to prevent overfitting under limited-data constraints [43].

$$\mu_y(k) \pm z_{1-\alpha/2}\sqrt{\sigma^2_{\text{tot},y}(k)} \qquad (23)$$

where $z_{0.975} = 1.96$ corresponds to the 95% intervals utilized in the experimental evaluation.

### *E. End-to-end training objective*

As shown in Panel E of Fig. 2, the MARS framework is optimized in an end-to-end differentiable manner. To ensure high prognostic accuracy while maintaining physical consistency, all modules are trained jointly using a multi-objective composite loss function:

$$\mathcal{L}_{\text{total}} = \mathcal{L}_{\text{RUL}} + \lambda_{\text{SOH}}\mathcal{L}_{\text{SOH}} + \lambda_{\text{phys}}\mathcal{L}_{\text{phys}} + \lambda_{\text{stable}}\mathcal{L}_{\text{stable}} \qquad (24)$$

where $\lambda_{\text{SOH}}$, $\lambda_{\text{phys}}$, and $\lambda_{\text{stable}}$ are weighting coefficients that balance the importance of health estimation, physical constraints, and filter stability.

For the primary tasks $y \in \{\text{RUL}, \text{SOH}\}$, we utilize a heteroscedastic Gaussian negative log-likelihood (NLL) to capture the cycle-dependent aleatoric uncertainty:

$$\mathcal{L}_{\text{NLL},y} = \frac{1}{2N}\sum_{k=1}^{N}\left[\frac{(y_k - \mu_{y,k})^2}{\sigma^2_{\text{ale},y,k} + \varepsilon} + \ln\left(\sigma^2_{\text{ale},y,k} + \varepsilon\right)\right] \qquad (25)$$

where $\mu_{y,k}$ and $\sigma^2_{\text{ale},y,k}$ denote the predicted mean and aleatoric variance, respectively, and $\varepsilon$ is a small constant for numerical stability.

To regularize non-physical model behavior, a physics-informed penalty $\mathcal{L}_{\text{phys}}$ is introduced. This term penalizes negative RUL predictions and non-physical increases in the capacity-related latent state, enforcing the monotonic nature of battery degradation:

$$\mathcal{L}_{\text{phys}} = \mathbb{E}\left[\text{ReLU}(-\mu_{\text{RUL},k})\right] + \mathbb{E}\left[\text{ReLU}(Q_{\max,k} - Q_{\max,k-1})\right] \qquad (26)$$

where $Q_{\max,k} \triangleq Q_{\max,k,L}$ represents the end-of-cycle capacity estimate derived from the GCAKF. Furthermore, to stabilize the recursive filtering process across diverse mini batches, we discourage excessive state dispersion using a stability loss:

$$\mathcal{L}_{\text{stable}} = \mathbb{E}\left[\|\hat{\boldsymbol{x}}_k - \mathbb{E}[\hat{\boldsymbol{x}}_k]\|_2^2\right] \qquad (27)$$

*1)Task-Driven Parameter Calibration:* A defining attribute of the proposed objective is the autonomous calibration of the GCAKF parameters $(Q, R, \sigma)$, which are treated as fixed hyperparameters in conventional EKF practice and are commonly tuned by residual inspection or trial-and-error. In contrast, MARS treats the full recursion as a sequence of differentiable operators. By backpropagating gradients of the task loss through the unrolled Kalman gain and state-update steps, the filter parameters are optimized in a task-driven manner under physical regularization. This yields a gain schedule that adapts to nonlinear degradation trajectories while preserving numerical stability and physical consistency.

The training and inference procedures of the proposed MARS framework are summarized in Algorithm 1 and Algorithm 2, respectively.

**Algorithm 1:** Training of MARS

**Require:** $\mathcal{D}$; $(L, W, \eta, E)$; $(\lambda_{\text{SOH}}, \lambda_{\text{phys}}, \lambda_{\text{stable}})$.

**Ensure:** $\Theta = \{\Theta_{\text{Enc}_s}, \Theta_{\text{Enc}_l}, \Theta_{\text{fus}}, \Theta_{\text{kf}}, \Theta_{\text{pred}}\}$.

**Initialization**
  Initialize $\Theta(\text{incl. } \Lambda_Q, \Lambda_R, (x_0, P_0))$.
  $\mathcal{O} \leftarrow \text{AdamW}(\Theta, \eta)$
**Optimization**
**for** $e \leftarrow 1$ **to** $E$ **do**
  **for all** mini-batch $\mathcal{B} \subset \mathcal{D}$ **do**
    **Inputs:**
      $X_k^{\text{short}}, X_k^{\text{long}} \leftarrow$ Eqs. (4)–(5); targets $\leftarrow$ Eqs. (1)–(3).
    **Encode & fuse**
      $(h_k^{(s)}, h_k^{(l)}) \leftarrow$ Eq. (6), (Eqs. (8)–(9)).
      $h_{f,k} \leftarrow$ Eq. (7)
    **GCAKF (differentiable):**
      $(Q_k, R_k) \leftarrow$ Eq. (13); $\sigma_k \leftarrow$ Eq. (14).
      $(\hat{x}_k, P_k) \leftarrow$ Eqs. (11)–(18) (GC: (15), infl.: (16), update: (17)).
    **Predict & loss**
      $z_k \leftarrow$ Eq. (19).
      $(\mu_{y,k}, s_{y,k}) \leftarrow g_y(z_k)$ for $y \in \{\text{RUL}, \text{SOH}\}$.
      $\sigma^2_{\text{ale},y,k} \leftarrow$ Eq. (20). (variance bounding)
      $\mathcal{L}_{\text{total}} \leftarrow$ Eq. (24)(NLL: (25), phys.: (26), stab.: (27))
      Enforce Eq. (26) on $\mu^{\text{norm}}_{\text{RUL},k} \in [0,1]$
    **Update**
      Backprop $\nabla_\Theta \mathcal{L}_{\text{total}}$; $\mathcal{O}$.step().
  **end for**
**end for**
**return** $\Theta$

**Algorithm 2:** Inference and uncertainty of MARS

**Require:** Trained $\Theta$; $(X_k^{short}, X_k^{long})$; $1-\alpha$; M.
**Ensure:** $\hat{\mu}_{y,k}$ and $(\hat{y}_k^-, \hat{y}_k^+)$, $y \in \{RUL, SOH\}$.
**Preprocessing and encoding:**
  $X_k^{short}, X_k^{long} \leftarrow$ Eqs. (4) – (5).
  $h_{f,k} \leftarrow$ Eqs. (6) – (9) and Eq. (7).
**Physics-consistent state estimation (GCAKF):**
  $(Q_k, R_k) \leftarrow$ Eq. (13); $\sigma_k \leftarrow$ Eq. (14).
  $(\hat{x}_k, P_k) \leftarrow$ Eqs. (11) – (18).
  $z_k \leftarrow$ Eq. (19).
MC-dropout sampling:
  Enable MC-dropout in predictor heads.
**for** $m \leftarrow 1$ **to** M **do**
    $(\mu_{y,k}^{(m)}, u_{y,k}^{(m)}) \leftarrow g_y(z_k)$ for $y \in \{RUL, SOH\}$
    $\sigma_{ale,y,k}^{2(m)} \leftarrow$ Eq. (20).
**end for**
**Uncertainty decomposition and interval:**
  $\hat{\mu}_{y,k} \leftarrow mean_m(\mu_{y,k}^{(m)})$.
  $\sigma_{epi,y,k}^2 \leftarrow Var_m(\mu_{y,k}^{(m)})$.
  $\sigma_{ale,y,k}^2 \leftarrow mean_m(\sigma_{ale,y,k}^{2(m)})$.
  $\sigma_{phys,k}^2 \leftarrow$ from $P_k$ and state-scale vector $\boldsymbol{s}$ by Eq. (22).
  $\sigma_{tot,y,k}^2 \leftarrow$ Eq. (21).
  $(\hat{y}_k^-, \hat{y}_k^+) \leftarrow$ Eq. (23).
**return** $\hat{\mu}_{y,k}$ and $(\hat{y}_k^-, \hat{y}_k^+)$ for $y \in \{RUL, SOH\}$

## IV. Results and Discussion

### A. Experimental settings and evaluation protocol

Experiments use the processed CALCE (CS2_35-CS2_38) and NASA (B0005, B0006, B0007, B0018) benchmarks constructed as described in Section II. To ensure a fair comparison, all methods share identical input definitions and splits. Features are min-max normalized using statistics computed on the training partition. Unless stated otherwise, we use a fixed seed (42) and a 3:1:1 train/validation/test split; key hyperparameters are reported in Table I. All models are implemented in Python (3.12) with PyTorch and evaluated on an NVIDIA GeForce RTX 4060 Laptop GPU (8GB) with CUDA.

We evaluate MARS against five representative baselines under identical experimental conditions: an Extended Kalman Filter combined with a Multilayer Perceptron (EKF-MLP), a Dual-Transformer, a Temporal Convolutional Network combined with a Gated Recurrent Unit (TCN-GRU), a parallel Extended Kalman Filter-Transformer architecture (EKF||Transformer), and a Physics-Informed Neural Network (PINN). In addition, FuseDR-Net [44] is discussed as a recent decomposition-based prognostic framework because it models capacity degradation and regeneration through separated temporal branches. Unlike FuseDR-Net, MARS does not explicitly decompose the capacity trajectory; instead, it couples multi-scale degradation representation with differentiable robust state estimation. Therefore, Table II reports only methods implemented under the same data definitions, inputs, and evaluation protocol. Component contributions are assessed using three ablations: without the GCAKF module, without the short-horizon branch, and without the long-horizon branch (Fig. 6 and Table III).

*1) Metrics and runtime.* For SOH and RUL point predictions, we report MAE, RMSE, and MAPE in Eq. (28), using $\varepsilon$ for numerical stability. SOH errors are reported in percentage points, and RUL errors in cycles. Uncertainty quality is evaluated using PICP and MPIW for nominal 95% intervals in Eq. (29). Computational performance is reported via inference latency/throughput and resource footprint measured on the same hardware:

$$\begin{aligned} \text{MAE} &= \frac{1}{N}\sum_{k=1}^{N}\left|y_k - \hat{y}_k\right|, \\ \text{RMSE} &= \sqrt{\frac{1}{N}\sum_{k=1}^{N}\left(y_k - \hat{y}_k\right)^2}, \\ \text{MAPE} &= \frac{100}{N}\sum_{k=1}^{N}\frac{\left|y_k - \hat{y}_k\right|}{\max\left(\left|y_k\right|, y_{min}\right)}, \end{aligned} \tag{28}$$

$$\begin{aligned} \text{PICP} &= \frac{100}{N}\sum_{i=1}^{N}\mathbb{I}\left(y_i \in [\hat{y}_i^-, \hat{y}_i^+]\right), \\ \text{MPIW} &= \frac{1}{N}\sum_{i=1}^{N}\left(\hat{y}_i^+ - \hat{y}_i^-\right), \end{aligned} \tag{29}$$

where $[\hat{y}_i^-, \hat{y}_i^+]$ denotes the predicted 95% interval.

*Table I Key Hyperparameters and GCAKF Initialization*

| Hyperparameter | Value |
|---|---|
| **Encoder and Prediction Modules** | |
| Short sequence length (*L*) | 50 |
| Long window size (*W*) | 20 cycles |
| Long feature dim ($D_{long}$) | 50 |
| Short Transformer ($d_s$, heads, layers) | (128, 8, 3) |
| Long Transformer ($d_\ell$, heads, layers) | (64, 2, 1) |
| Fusion Dim / Prediction hidden | 128 / 128 |
| Dropout | 0.0899 |
| Optimizer / LR / Weight decay | AdamW / $10^{-4}$ / $10^{-4}$ |
| Batch size / Epochs/patience | 64 / 200 / 20 |
| Grad clip | 0.5 |
| Loss weight ($\lambda_{SOH}, \lambda_{phys}, \lambda_{stable}$) | (0.5, 3e-4, 1e-4) |
| **GCAKF Initialization** | |
| Initial state $\boldsymbol{x}_0$ | $[1.0, 0, 1.1, 0.1, 0.05, 1000]^T$ |
| Initial covariance $\boldsymbol{P}_0$ | $diag([10^{-3}, 10^{-3}, 5\times10^{-2}, 5\times10^{-2}, 5\times10^{-2}, 10^{-1}])$ |
| Process noise $\boldsymbol{Q}$ | $diag([10^{-6}, 10^{-5}, 5\times10^{-5}, 10^{-4}, 10^{-4}, 5\times10^{-4}])$ |
| Measurement noise *R* | $10^{-4}$ |

### B. SOH Estimation Accuracy and Cross-Regime Consistency

The first evaluation examines whether MARS preserves accurate SOH tracking across different degradation horizons and cell-aging patterns. The NASA and CALCE results are presented separately in Figs. 4(a)-(l) and 5(a)-(l), respectively. For each dataset, panels (a)-(d) show the estimated SOH trajectories, panels (e)-(h) present the signed residuals, and panels (i)-(l) compare the cell-level MAE, RMSE, and MAPE.

The NASA cells exhibit markedly different degradation rates and EOL timings. B0006 degrades more rapidly, whereas B0005, B0007, and B0018 follow slower and nonuniform trajectories. As shown in Fig. 4(a)-(h), MARS follows these cell-specific trends without the persistent offset or trajectory flattening observed for several baselines. Its residuals remain concentrated around zero, indicating reduced systematic bias and local estimation error. Table II confirms this behavior: MARS achieves an SOH MAE of 0.72%, RMSE of 0.92%, and MAPE of 0.87%. Relative to PINN, the strongest competing SOH estimator, these values correspond to reductions of 36.3%, 39.9%, and 34.1%, respectively.

CALCE provides a longer degradation horizon of approximately 700 cycles, including gradual capacity loss followed by a pronounced late-life knee. Fig. 5(a)-(h) shows that MARS tracks both regions while avoiding progressive residual drift. It obtains an SOH MAE of 0.63%, RMSE of 0.80%, and MAPE of 1.00%, compared with 1.10%, 1.48%, and 1.72% for EKF||Transformer. The lower RMSE is particularly important because it indicates fewer large deviations around the nonlinear degradation knee.

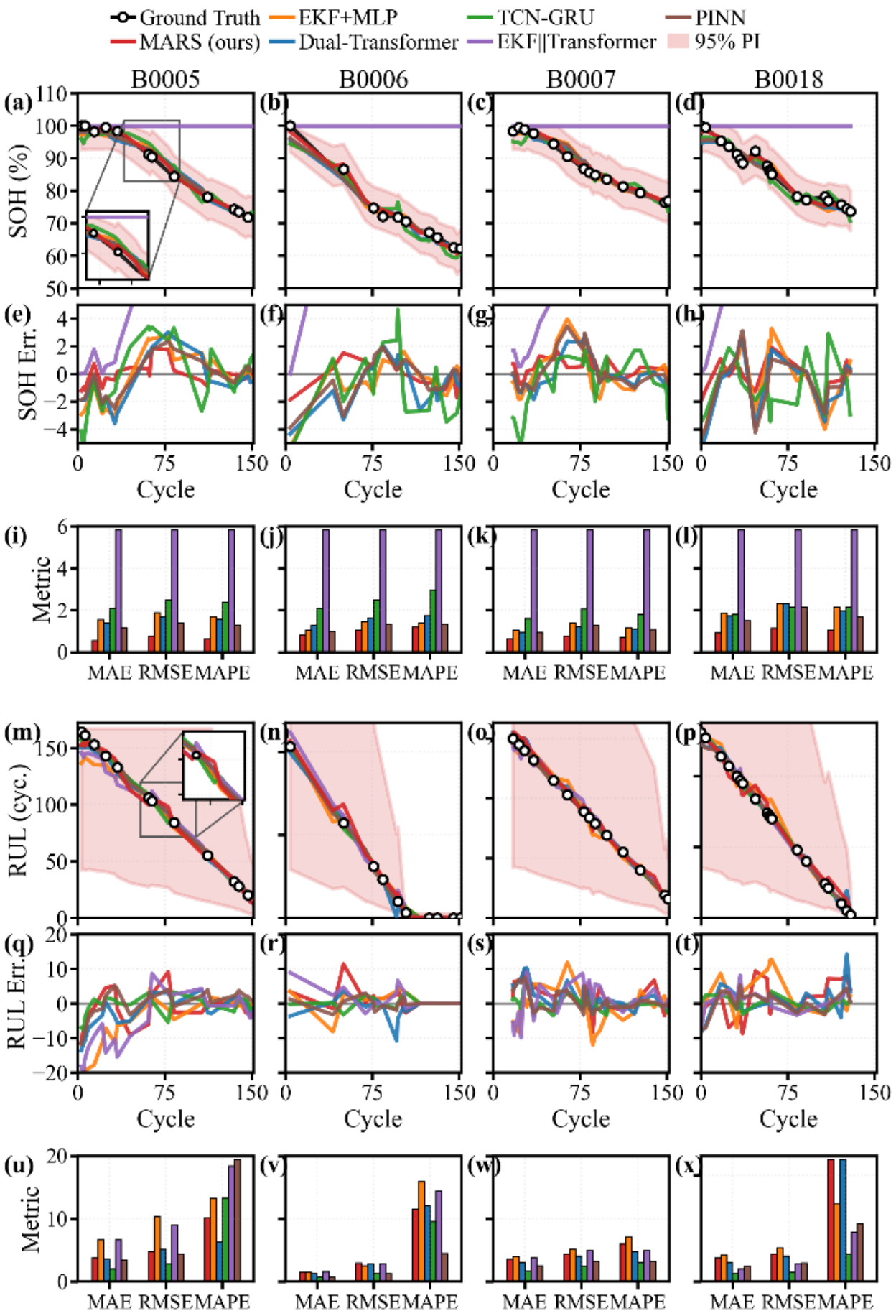

*Fig. 4. SOH and RUL results for NASA cells B0005, B0006, B0007, and B0018: (a)-(l) SOH tracking and errors; (m)-(x) RUL tracking, 95% PIs, and errors.*

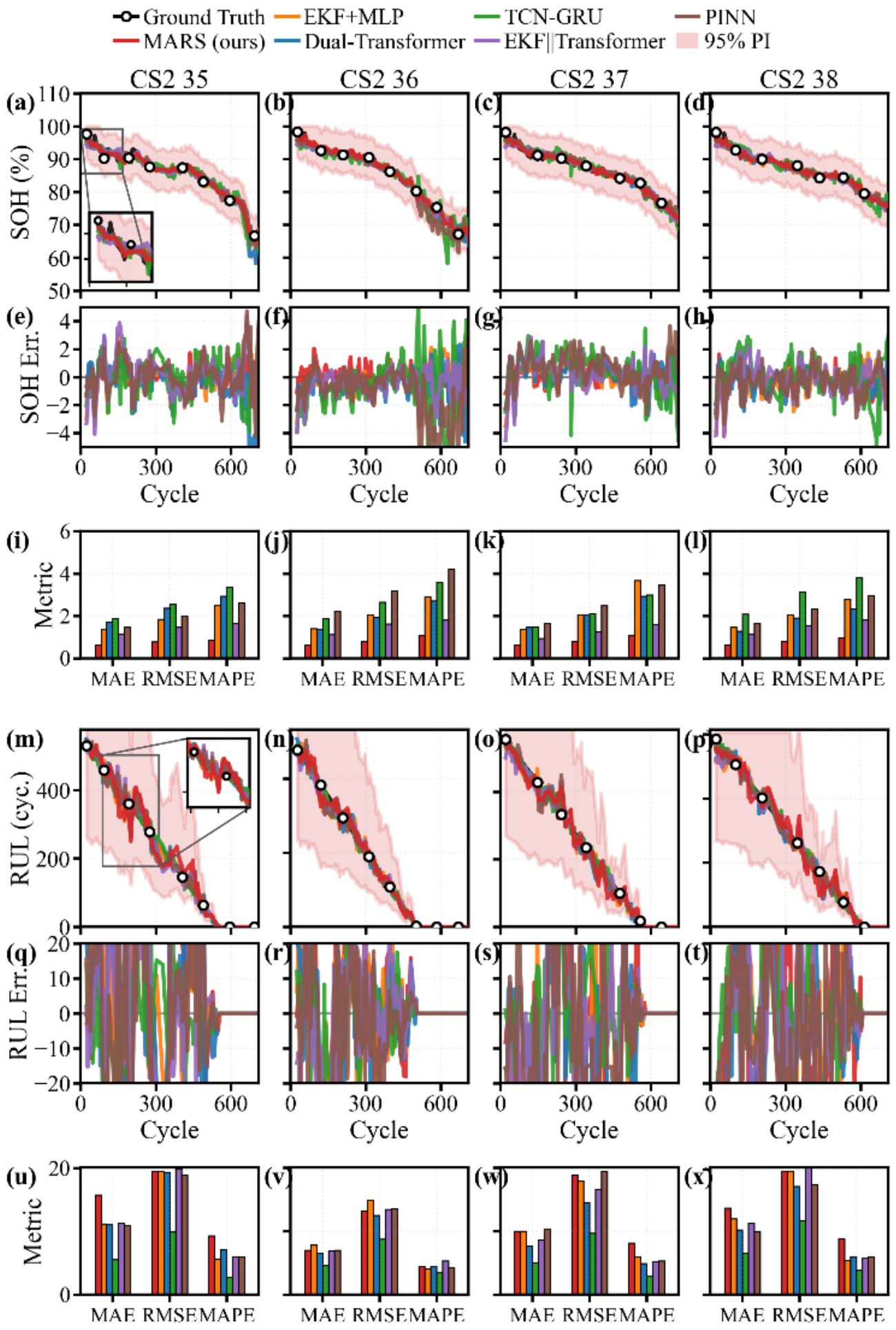

*Fig. 5. SOH and RUL results for CALCE cells CS2-35-CS2-38: (a)-(l) SOH tracking and errors; (m)-(x) RUL tracking, 95% PIs, and errors.*

This cross-regime stability is consistent with the MARS formulation. The short-horizon encoder captures intra-cycle voltage, current, and temperature dynamics, whereas the long-horizon encoder retains the slower inter-cycle aging context. Their fused representation adapts $\boldsymbol{Q}_k$ and $\boldsymbol{R}_k$ through Eq. (13) and the robustness bandwidth through Eq. (14). Large innovations are then attenuated by the generalized-correntropy weight in Eq. (15), followed by the covariance and posterior updates in Eqs. (16)-(18).

### C. RUL prognostics and uncertainty behavior

The second evaluation examines RUL prognostics, which differ from SOH estimation because RUL is conditional on future operation remaining consistent with the observed degradation history. The NASA and CALCE results are shown in Figs. 4(m)-(x) and 5(m)-(x), respectively. Panels (m)-(p) present the RUL trajectories with 95% prediction intervals, panels (q)-(t) show the signed errors, and panels (u)-(x) report cell-level MAE, RMSE, and MAPE. Table II summarizes the corresponding point accuracy and interval quality.

On NASA, MARS achieves an MAE of 3.32 cycles, RMSE of 4.44 cycles, and MAPE of 12.91%, with 100% PICP and a mean PI width of 101.77 cycles. TCN-GRU yields lower point errors, whereas Dual-Transformer under-covers the observations with a PICP of 92.71%. The relatively high NASA

MAPE mainly reflects the amplification of percentage errors as the true RUL approaches zero.

On CALCE, the uncertain timing of the degradation knee increases long-horizon difficulty. MARS obtains an MAE of 11.61 cycles, RMSE of 21.34 cycles, and MAPE of 7.65%, while maintaining 99.24% coverage with a mean PI width of 161.80 cycles. Although TCN-GRU provides lower point errors, its PICP falls to 93.50%. The intervals are derived from Eqs. (21)-(23), including the GCAKF state-related uncertainty contribution.

### *D. Ablation study*

Table III and Fig. 6 isolate the contributions of the GCAKF, the short-horizon branch, and the long-horizon branch. The most severe degradation occurs when the long-horizon branch is removed. On CALCE, the RUL MAE increases from 11.61 to 66.68 cycles and the SOH MAE from 0.63% to 4.32%; on NASA, the corresponding errors increase from 3.32 to 50.01 cycles and from 0.72% to 10.35%. This confirms that long-range degradation context is essential for preserving the slowly evolving aging state and preventing error accumulation across cycles.

Removing the short-horizon branch produces a smaller but still meaningful degradation. On NASA, both SOH and RUL errors increase, indicating that intra-cycle voltage, current, and temperature signatures are important for tracking heterogeneous local dynamics. On CALCE, the RUL point error changes less markedly, but SOH accuracy still deteriorates, showing that the short-horizon branch mainly strengthens local state tracking rather than replacing the long-term degradation context.

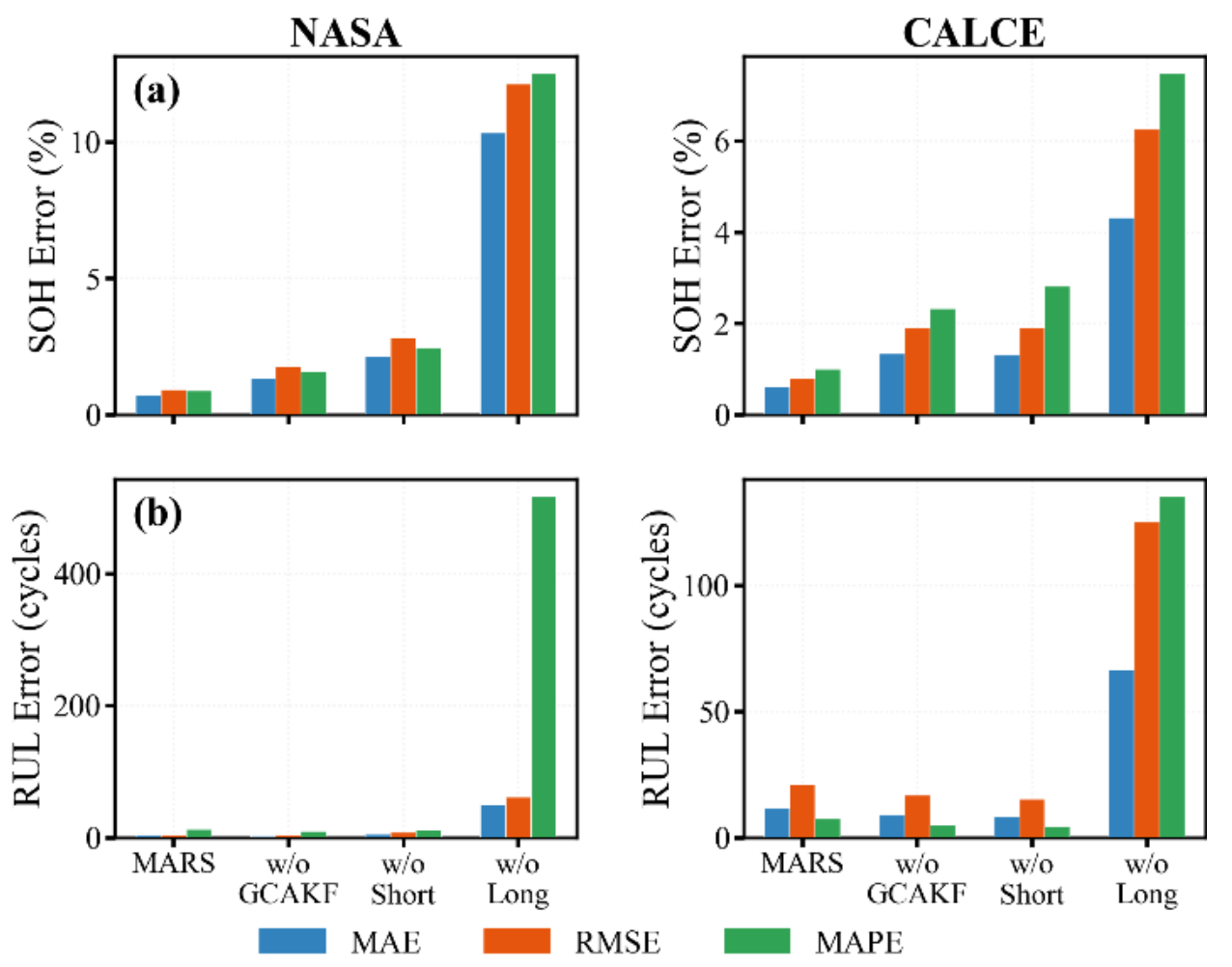

*Fig. 6. Ablation results of MARS on NASA and CALCE: (a) SOH error metrics and (b) RUL error metrics for the full model.*

Removing the GCAKF has a different effect. The point errors change less dramatically than in the long-branch ablation, but the uncertainty quality degrades clearly. On CALCE, the PICP decreases from 99.24% to 96.69% while the mean PI width contracts from 161.80 to 48.47 cycles; On NASA, the PICP decreases from 100.00% to 91.67%, while the mean PI width contracts from 101.77 to 16.57 cycles. Thus, the GCAKF is not only a state-correction module, but also the main source of robust innovation weighting and calibrated uncertainty propagation. Overall, the ablation results show that the long-horizon branch provides the dominant degradation memory, the short-horizon branch improves local tracking, and the GCAKF ensures physically consistent and reliable prediction.

Table II SOH/RUL performance on CALCE (CS2-35-CS2-38) and on NASA (B0005/B0006/B0007/B0018) SOH (%) and RUL (cycle).

| Dataset | Method | RUL MAE | RUL RMSE | RUL MAPE | SOH MAE | SOH RMSE | SOH MAPE | PICP | 95% PI Width (cycle) |
|---|---|---|---|---|---|---|---|---|---|
| CALCE | **MARS (Ours)** | **11.61** | **21.34** | **7.65** | **0.63** | **0.80** | **1.00** | **99.24** | **161.80** |
| | Dual-Transformer | 8.90 | 16.06 | 5.01 | 1.44 | 2.07 | 2.71 | 97.20 | 56.84 |
| | TCN-GRU | 5.44 | 10.16 | 3.32 | 1.84 | 2.65 | 3.46 | 93.50 | 23.57 |
| | EKF-MLP | 10.28 | 18.43 | 5.29 | 1.41 | 2.00 | 2.96 | 100.00 | 160.64 |
| | EKF\|\|Transformer | 9.59 | 17.76 | 5.54 | 1.10 | 1.48 | 1.72 | 99.87 | 162.04 |
| | PINN | 9.56 | 17.68 | 5.42 | 1.74 | 2.54 | 3.32 | 96.05 | 47.93 |
| NASA | **MARS (Ours)** | **3.32** | **4.44** | **12.91** | **0.72** | **0.92** | **0.87** | **100.00** | **101.77** |
| | Dual-Transformer | 2.94 | 4.31 | 11.87 | 1.31 | 1.70 | 1.55 | 92.71 | 19.13 |
| | TCN-GRU | 1.53 | 2.24 | 7.74 | 1.90 | 2.31 | 2.30 | 96.88 | 12.57 |
| | EKF-MLP | 4.39 | 6.84 | 12.31 | 1.34 | 1.76 | 1.56 | 100.00 | 101.94 |
| | EKF\|\|Transformer | 3.83 | 5.79 | 11.66 | 17.10 | 20.79 | 23.23 | 98.96 | 101.52 |
| | PINN | 2.45 | 3.37 | 11.42 | 1.13 | 1.53 | 1.32 | 95.83 | 12.23 |

Table III Ablation summary of SOH/RUL performance on CALCE and NASA. SOH is reported in % and RUL in cycles.

| Dataset | Method | RUL MAE | RUL RMSE | RUL MAPE | SOH MAE | SOH RMSE | SOH MAPE | PICP | 95% PI Width (cycles) |
|---|---|---|---|---|---|---|---|---|---|
| CALCE | **MARS (Full)** | **11.61** | **21.34** | **7.65** | **0.63** | **0.80** | **1.00** | **99.24** | **161.80** |
| | w/o GCAKF | 8.97 | 16.97 | 5.11 | 1.34 | 1.90 | 2.32 | 96.69 | 48.47 |
| | w/o short branch | 8.38 | 15.45 | 4.72 | 1.32 | 1.91 | 2.82 | 100.00 | 193.08 |
| | w/o long branch | 66.68 | 125.11 | 135.33 | 4.32 | 6.26 | 7.48 | 94.39 | 283.26 |
| NASA | **MARS (Full)** | **3.32** | **4.44** | **12.91** | **0.72** | **0.92** | **0.87** | **100.00** | **101.77** |
| | w/o GCAKF | 2.76 | 4.30 | 10.56 | 1.34 | 1.80 | 1.59 | 91.67 | 16.57 |
| | w/o short branch | 5.50 | 8.29 | 12.19 | 2.15 | 2.84 | 2.47 | 98.96 | 61.14 |
| | w/o long branch | 50.01 | 61.83 | 517.49 | 10.35 | 12.13 | 12.52 | 100.00 | 167.00 |

### E. Lifecycle-Resolved Uncertainty and Near-EOL Behavior

Lifecycle-dependent confidence is important in electrified transport because the earliest-degrading cell can constrain the usable capacity and power capability of the complete pack. Fig. 9 and Table IV therefore evaluate MARS from early life to the final 30 cycles before EOL using the uncertainty formulation in Eqs. (20)-(23).

On NASA, SOH MAE remains below 1% in every stage, while RUL MAE decreases from 5.19 cycles in early life to 2.00 cycles near EOL. The 95% PI width increases to 145.43 cycles in middle life and then contracts to 39.82 cycles near EOL, with 100% coverage throughout. The broad intermediate-life intervals reflect uncertainty in the still-unobserved degradation path, whereas the shorter remaining horizon provides stronger evidence of EOL as aging progresses.

The decomposition in Fig. 9(i)-(l) follows Eqs. (20)-(22). Aleatoric uncertainty represents observation-dependent variability, epistemic uncertainty is estimated through MC dropout, and the physics-induced term propagates the GCAKF posterior covariance $\boldsymbol{P}_k$ through Eq. (22). On NASA, the physics-induced fraction decreases from 98.84% in early life to 84.20% near EOL, while the aleatoric fraction rises to 12.74%, indicating increasing influence of local degradation and measurement variability.

CALCE exhibits a stronger horizon and knee-point effect. Its PI width peaks at 431.19 cycles in early-middle life, when the future nonlinear decline remains difficult to locate. Aleatoric and epistemic contributions increase during the middle and late-middle stages as the degradation knee develops. Near EOL, RUL MAE falls to 0.38 cycles and the PI width contracts to 2.99 cycles while maintaining 98.72% coverage. Although the physics-induced fraction reaches 96.22%, the narrow interval shows that the absolute uncertainty is small; most of the remaining variance is associated with state-covariance propagation.

Thus, MARS provides broad bounds when long-term degradation remains uncertain and tighter bounds when EOL becomes observable. This behavior can support cell inspection, maintenance, pack retirement, and preliminary second-life screening without relying on a single deterministic RUL estimate.

### F. Sensitivity to Covariance Prior and Non-Gaussian Disturbances

An adaptive state estimator should remain stable without requiring a precisely tuned initial covariance, since the actual confidence in the battery state is generally unknown at deployment. We therefore vary the covariance-prior scale over $\boldsymbol{\theta} \in \{0.1, 1, 10\}$ to examine whether MARS is sensitive to this initialization. As shown in Fig. 7(a) and (c), the SOH and RUL errors change only moderately on both CALCE and NASA. The uncertainty results in Fig. 7(b) and (d) show that $\boldsymbol{\theta} = 0.1$ can produce lower coverage, particularly on CALCE, whereas $\boldsymbol{\theta} = 1$ provides a better balance between point accuracy and interval reliability. Increasing the scale to $\boldsymbol{\theta} = 10$ provides no clear benefit and slightly increases the RUL error. The limited dependence on $\boldsymbol{\theta}$ is consistent with Eq. (13), because the context-conditioned $\boldsymbol{Q}_k$ and $\boldsymbol{R}_k$ continuously adjust the relative confidence assigned to state propagation and incoming measurements.

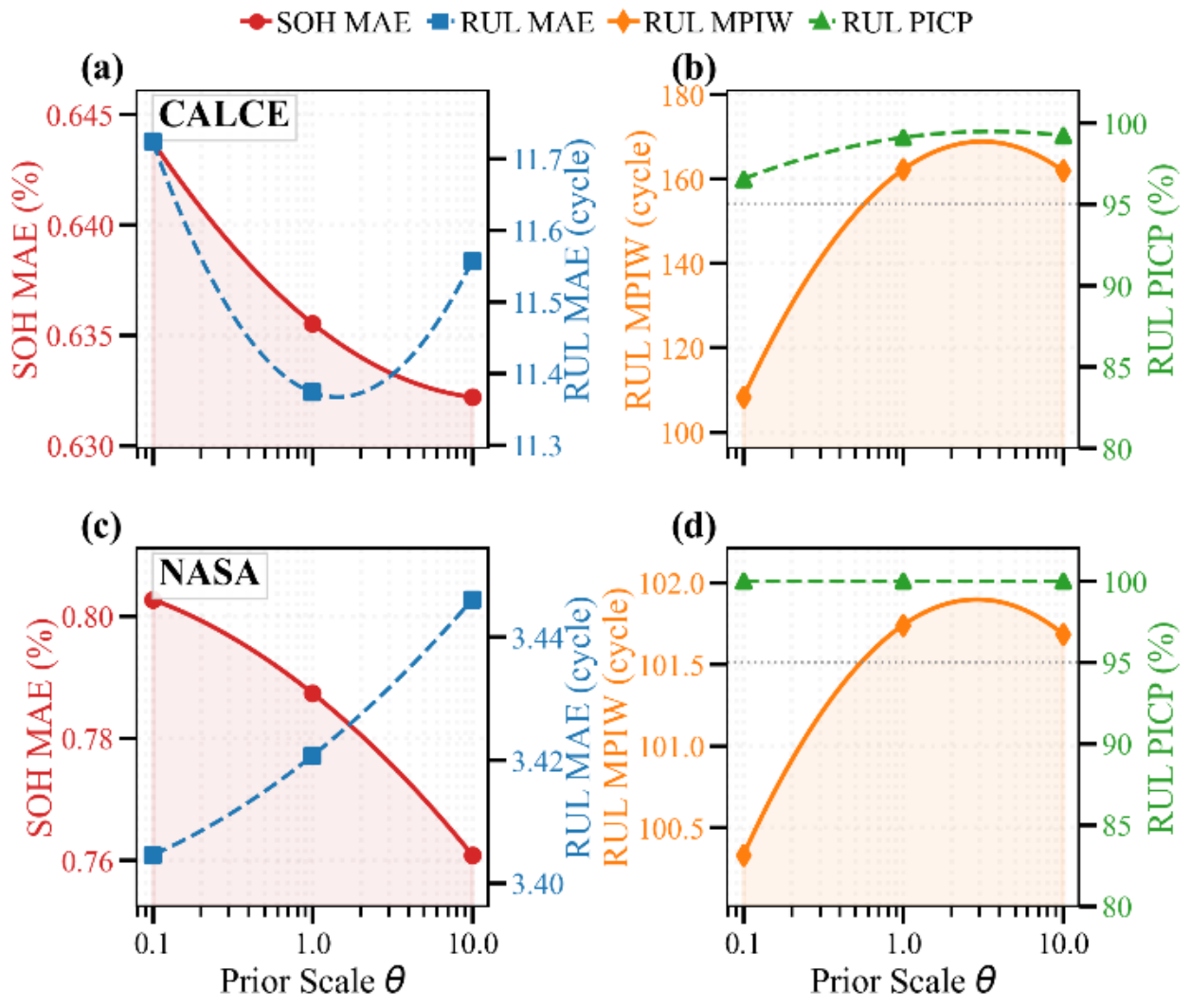

*Fig. 7. Sensitivity of MARS to the covariance-prior scale $\boldsymbol{\theta}$: (a) CALCE SOH and RUL MAE, (b) CALCE RUL MPIW and PICP, (c) NASA SOH and RUL MAE, and (d) NASA RUL MPIW and PICP.*

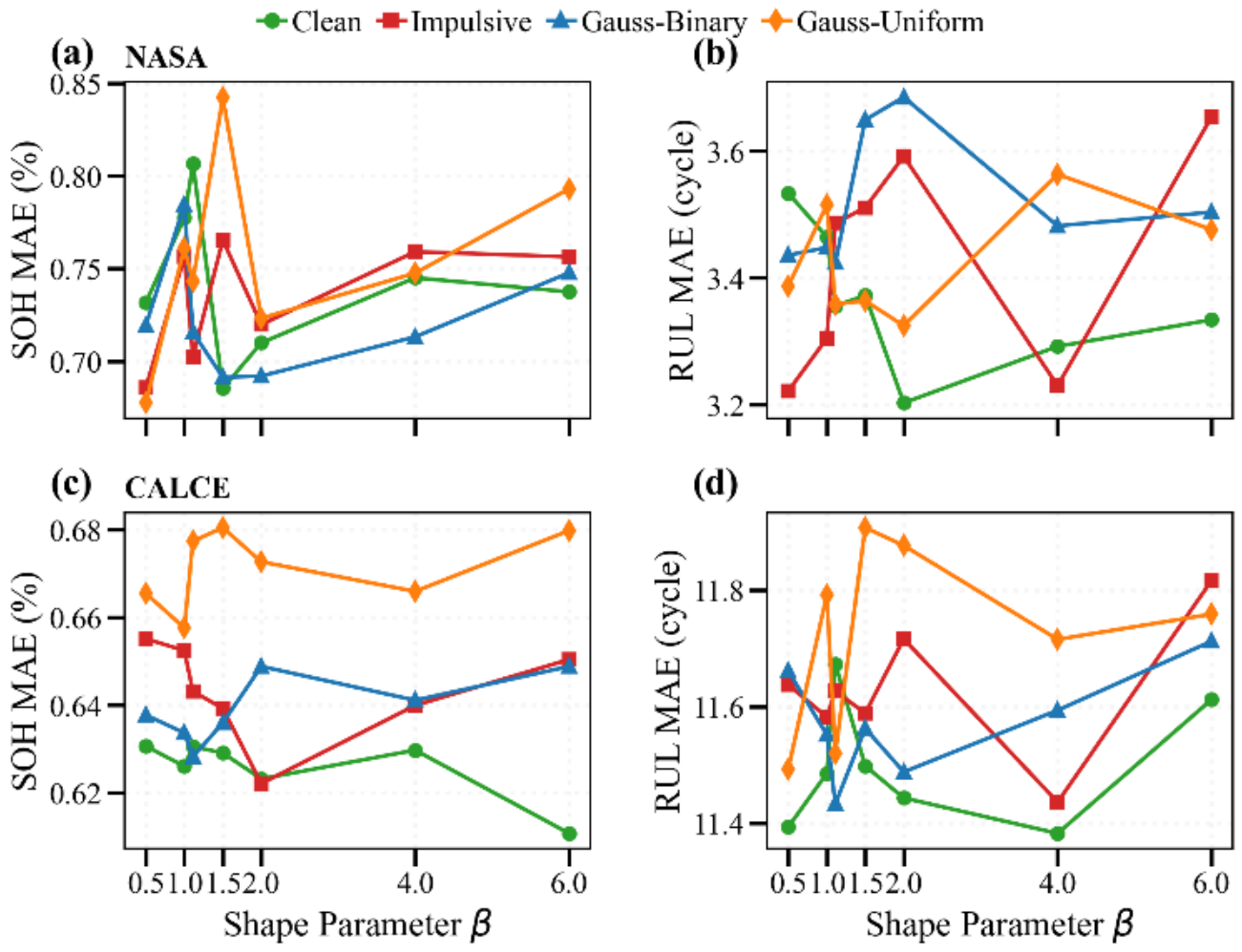

Fig. 8. Sensitivity of MARS to the generalized-correntropy shape parameter $\boldsymbol{\beta}$ under clean and non-Gaussian disturbances: (a) NASA SOH MAE, (b) NASA RUL MAE, (c) CALCE SOH MAE, and (d) CALCE RUL MAE.

Battery measurements may also contain impulsive spikes and mixed non-Gaussian disturbances that violate the Gaussian-noise assumption of conventional Kalman filtering. To evaluate this robustness, the generalized-correntropy shape parameter $\boldsymbol{\beta}$ is varied under clean, impulsive, Gaussian-binary, and Gaussian-uniform conditions. Fig. 8 shows that the SOH and RUL errors remain bounded over the tested range on both datasets. On NASA, SOH MAE remains approximately within 0.67-0.84% and RUL MAE within 3.20-3.70 cycles; on CALCE, the corresponding ranges are approximately 0.61-0.68% and 11.4-12.0 cycles. No single extreme value of $\boldsymbol{\beta}$ consistently minimizes all errors, while intermediate values around $\boldsymbol{\beta} = 1-2$ provide the most balanced response across disturbance types. This behavior agrees with Eq. (15): smaller $\boldsymbol{\beta}$ suppresses large innovations more strongly, whereas larger $\beta$ reduces the distinction between nominal residuals and outliers.

These tests indicate that MARS robustness is produced by adaptive covariance correction and generalized-correntropy innovation weighting, rather than by narrowly tuned prior or shape-parameter values.

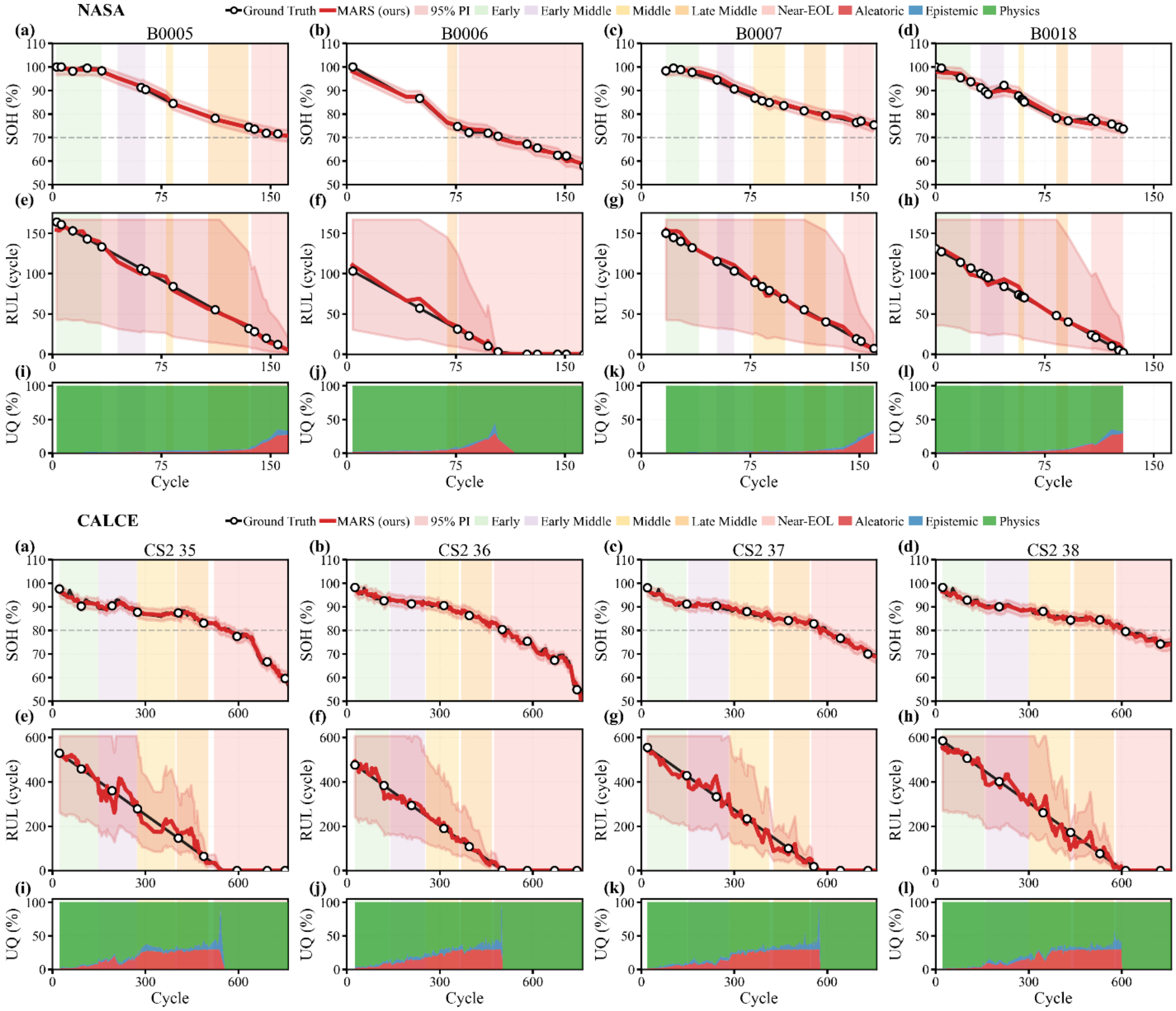

*Fig. 9. Lifecycle-resolved SOH, RUL, and uncertainty results on NASA (top) and CALCE (bottom): (a)-(d) SOH trajectories, (e)-(h) RUL trajectories with 95% prediction intervals, and (i)-(l) aleatoric, epistemic, and physics-induced uncertainty contributions across the aging stages and near-EOL region.*

Table IV Lifecycle-resolved SOH MAE (%), RUL MAE and MPIW (cycles), PICP (%), and uncertainty contributions for NASA and CALCE.

| Aging Stage | SOH MAE | RUL MAE | RUL PICP | RUL MPIW | Aleatoric (%) | Epistemic (%) | Physics (%) |
|---|---|---|---|---|---|---|---|
| CALCE | | | | | | | |
| Early stage: 0-25% | 0.685 | 17.7 | 100 | 370.66 | 3.3 | 1.6 | 95.2 |
| Early middle stage: 25-50% | 0.542 | 29.1 | 100 | 431.19 | 11.3 | 3.6 | 85.1 |
| Middle stage: 50-75% | 0.561 | 23 | 100 | 327.68 | 23.7 | 5.1 | 71.3 |
| Late stage: 75-100% | 0.485 | 21.5 | 98 | 149.21 | 28.8 | 7 | 64.2 |
| Near-EOL: final 30 cycles | 0.677 | 0.4 | 98.7 | 2.99 | 2.1 | 1.7 | 96.2 |
| NASA | | | | | | | |
| Early stage: 0-25% | 0.860 | 5.2 | 100 | 128.18 | 0.7 | 0.4 | 98.8 |
| Early middle stage: 25-50% | 0.914 | 4.6 | 100 | 139.70 | 1.4 | 1.1 | 97.5 |
| Middle stage: 50-75% | 0.839 | 4.7 | 100 | 145.43 | 1.8 | 1.4 | 96.9 |
| Late stage: 75-100% | 0.313 | 2.1 | 100 | 141.60 | 3.8 | 1.6 | 94.6 |
| Near-EOL stage: final 30 cycles | 0.593 | 2 | 100 | 39.82 | 12.7 | 3.1 | 84.2 |

### G. Computational efficiency and deployment considerations

Practical BMS deployment requires a balance among estimation accuracy, uncertainty reliability, and computational cost. Fig. 10 compares all methods on the same NVIDIA RTX 4060 Laptop GPU using latency, throughput, Monte Carlo latency, uncertainty-quantification latency, parameter count, FP32 model size, peak inference memory, and energy per inference. NASA and CALCE results are shown in Fig. 10(a) and (b), respectively. Latency was measured after GPU warm-up using CUDA events, while peak inference memory was recorded using the PyTorch CUDA memory profiler. Because these quantities depend on hardware and implementation, they are interpreted as relative comparisons under identical benchmarking conditions, consistent with previous battery-learning and error-effort studies [7], [32], [40].

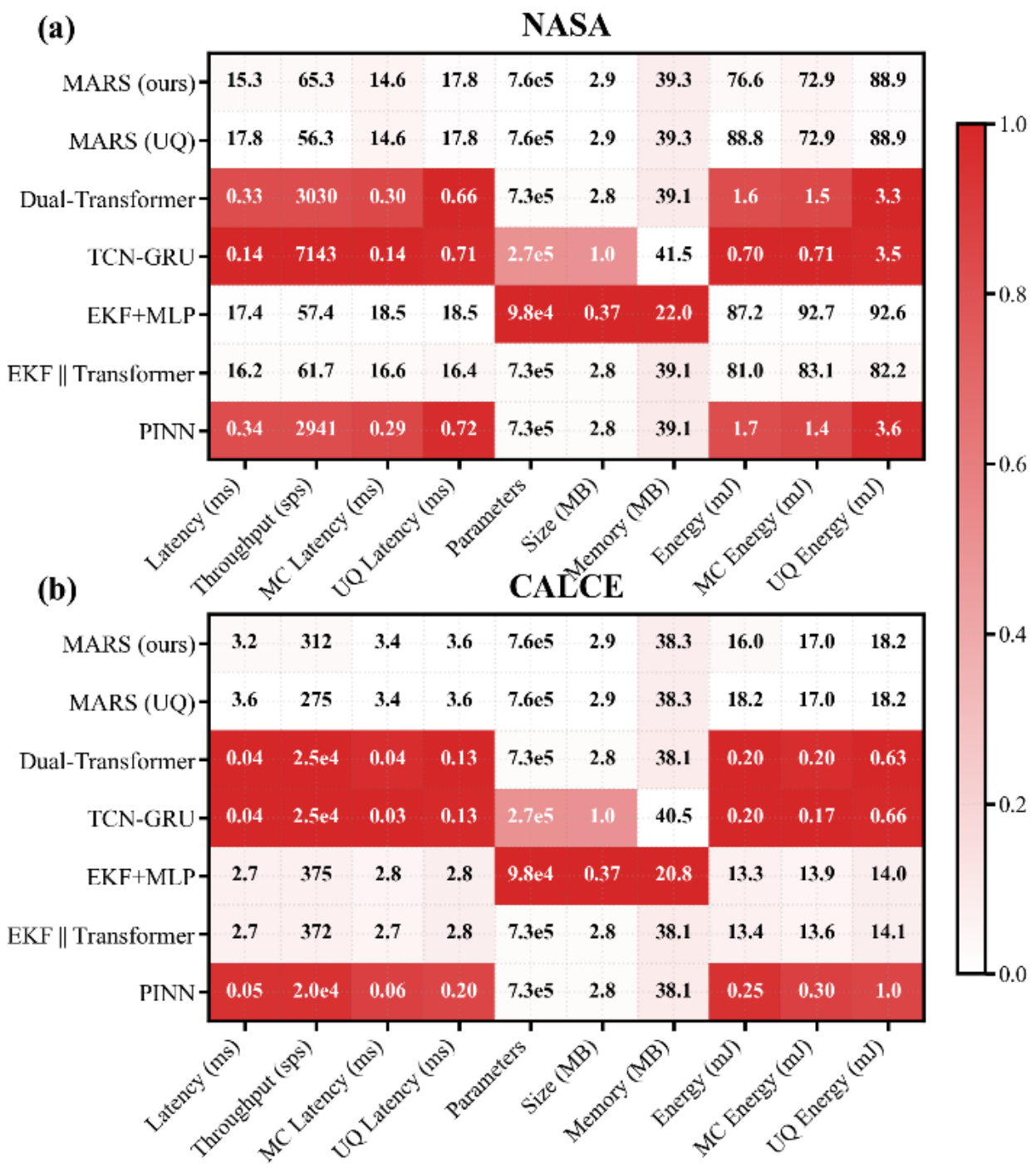

*Fig. 10. Computational profiles of MARS and the baseline methods on (a) NASA and (b) CALCE, including latency, throughput, parameter count, model size, peak inference memory, and energy per inference.*

The measured inference-time GPU power was approximately 3-4 W. To avoid underestimating consumption, the energy per inference was conservatively estimated as

$$E_{inf} = P_{env} t_{inf}, \qquad P_{env} = 5\,W,$$

where $t_{inf}$ is the measured inference latency. The reported values therefore represent conservative energy estimates on the evaluated GPU platform.

On CALCE, deterministic and uncertainty-aware MARS require 3.20 and 3.64 ms/sample, corresponding to 312 and 275 samples/s and 16.0 and 18.2 mJ/sample, respectively. On NASA, the corresponding values are 15.3 and 17.8ms/sample, 65.3 and 56.3 samples/s, and 76.6 and 88.9 mJ/sample. Uncertainty quantification therefore introduces only moderate additional overhead. Learning-only models remain faster, whereas MARS is computationally comparable to the EKF-based hybrids.

MARS contains 758,928 trainable parameters, requiring approximately 2.90MB of FP32 storage, with peak inference memory of approximately 38-39MB. These results establish cell-level computational feasibility. Following the error-effort principle in [19], deterministic inference may operate continuously, while uncertainty-aware inference is activated periodically or before maintenance and replacement decisions.

## V. Conclusion

This paper presented MARS, a multi-scale adaptive robust framework for SOH estimation and conditional RUL prediction with quantified uncertainty. The framework couples a dual-branch temporal encoder with a differentiable generalized-correntropy adaptive Kalman filter, providing physics-consistent state propagation and inherent robustness to non-Gaussian disturbances. Evaluations on the NASA and CALCE benchmarks demonstrate that MARS achieves high SOH accuracy and competitive conditional RUL performance while maintaining well-calibrated 95% prediction intervals across diverse degradation regimes. Ablation studies confirm that the long-horizon degradation context and the robust filtering module are essential for both point accuracy and interval reliability, and sensitivity analyses verify stable operation under covariance-prior variation and mixed non-Gaussian noise. Computational profiling on a laptop GPU shows inference times of a few milliseconds per cycle, supporting online battery management feasibility. Future work will extend the framework to pack-level prognostics with thermal and cell-to-cell coupling and will investigate energy-aware compression and edge-optimized deployment for real-time BMS integration.

## Conflict of Interest

The authors declare that they have no conflict of interest.